\documentclass[reprint,linenumber, amsmath amssymb, titlepage, aps, pra, superscriptaddress, raggedbottom, nofootinbib, floatfix, citeautoscript]{revtex4-2}
\usepackage{silence}
\usepackage[utf8]{inputenc}
\usepackage[english]{babel}
\usepackage{hyperref}
\usepackage{bm}
\usepackage{hyperref}
\usepackage{float}
\usepackage[newcommands]{ragged2e}
\usepackage{physics}
\usepackage{tabularx}
\usepackage{mathtools}
\usepackage[raggedright]{titlesec}
\usepackage[toc,page,header]{appendix}
\usepackage{minitoc}
\usepackage{nicefrac}
\usepackage{sidecap}
\usepackage{multibib}
\usepackage[dvipsnames]{xcolor}
\usepackage{booktabs}
\usepackage{float}
\usepackage{graphicx}
\usepackage[capitalise]{cleveref}
\usepackage[version=4]{mhchem}
\usepackage{tabularx}
\usepackage{ulem}
\usepackage{enumitem}

\sidecaptionvpos{figure}{c}
\usepackage[per-mode=power, 
            exponent-product = \times, 
            inter-unit-product = \ensuremath{{\cdot}}, 
            tight-spacing = true,
            number-unit-product=~,
            parse-numbers=false,]{siunitx}
\AtBeginDocument{\RenewCommandCopy\qty\SI}
\ExplSyntaxOn
\msg_redirect_name:nnn { siunitx } { physics-pkg } { none }
\ExplSyntaxOff
\hypersetup{colorlinks=true,  linkcolor=blue, 
            citecolor=blue, filecolor=blue,  
            urlcolor=blue}
\usepackage{caption}
\titleformat*{\section}{\bfseries}
\titleformat*{\subsection}{\bfseries}
\titlespacing{\section}{0em}{1em}{0.25em}
\renewcommand{\textcite}[1]{\citenum{#1}}
\newcolumntype{Y}{>{\centering\arraybackslash}X}
\Crefformat{appendix}{Supplementary Information #2#1#3}
\let\origaddcontentsline\addcontentsline

\newcommand{\enabletocentries}{%
  \let\addcontentsline\origaddcontentsline
}
\setcitestyle{super}
\DeclareSIUnit[qualifier-mode = combine]{\dBm}{\deci\bel\of{m}}

\makeatletter
\def\blfootnote{\gdef\@thefnmark{}\@footnotetext}
\makeatother

\begin{document}

\author{Shao-Chien Ou}
\email{sou12@umd.edu}
\affiliation{Joint Quantum Institute, NIST/University of Maryland, College Park, USA}
\affiliation{Microsystems and Nanotechnology Division, National Institute of Standards and Technology, Gaithersburg, USA}
\author{Gr\'egory~Moille}
\affiliation{Joint Quantum Institute, NIST/University of Maryland, College Park, USA}
\affiliation{Microsystems and Nanotechnology Division, National Institute of Standards and Technology, Gaithersburg, USA}
\author{Kartik~Srinivasan}
\email{kartik.srinivasan@nist.gov}
\affiliation{Joint Quantum Institute, NIST/University of Maryland, College Park, USA}
\affiliation{Microsystems and Nanotechnology Division, National Institute of Standards and Technology, Gaithersburg, USA}
\date{\today}

\newcommand{\mytitle}{Broadband Chromatic Dispersion of Thermo-refractive Coefficients and its Impact in Silicon Nitride Nonlinear Photonics
}
\title{\mytitle}

\begin{abstract}
  The thermo-refractive effect is a cornerstone of frequency and phase tuning in photonic integrated circuits. In particular, it enables control of phase-matching for integrated nonlinear processes. 
  Chromatic dispersion of the group and effective refractive indices and modal confinement are standard considerations in design, but material thermo-refractive coefficients (TRCs) are typically taken to be fixed for the guiding and cladding materials.  
  Here, we demonstrate that the assumption of non-dispersive TRCs across an octave of bandwidth between the telecom and visible results in a significant discrepancy between measured and simulated resonance frequencies of an integrated \ce{Si3N4}/\ce{SiO2} microring resonator.
  We uncover a \qty{\approx 7}{\percent} variation in \ce{Si3N4} and \ce{SiO2} material TRCs across this range, finding that the variation of $\mathrm{d}n_\mathrm{eff}/\mathrm{d}T$ from material TRCs is 1.3 times that from modal confinement. %
  This accurately matches a temperature-dependent Lorentz oscillator model describing their chromatic dispersion.  
  By integrating these dispersive TRCs into a multi-physics finite-element model, we achieve precise correspondence with experimentally measured temperature-dependent resonance frequency shifts across the octave, including in the context of second harmonic generation devices. Our results provide a physical framework and a universal predictive workflow for the design of high-efficiency, multi-wavelength nonlinear optical processes, fundamentally improving the thermal control of integrated photonic devices.

  
\end{abstract}
\maketitle

\section{Introduction}

\blfootnote{$^{\ddagger}$~This document is preliminary and is intended for peer review conducted by a journal.}

  The generation of new optical frequencies fundamentally relies on the energy and momentum conservation imposed by the nonlinear interactions~\cite{clementi_ultrabroadband_2025}, with applications in optical metrology~\cite{newman_architecture_2019}, biological imaging~\cite{campagnola_second_2011}, and high-speed communications~\cite{geng_terabit_2018}. Photonic integration has enabled significant reduction in device footprint~\cite{delhaye_optical_2007}, while the advent of high-quality-factor on-chip resonators has enabled reduction in power consumption~\cite{ferrera_low_2009}. However, fabrication tolerances generally do not achieve the precision required to produce experimental microresonator systems that provide perfect frequency and phase matching in as-fabricated devices~\cite{ferraro_imec_2023}. Therefore, post-fabrication tuning remains a critical tool for achieving resonance alignment, and is commonly realized through the thermo-refractive effect~\cite{bosman_temperature_1963}, which enables refractive index tuning via heat injection~\cite{liu_thermo-optic_2022, parra_silicon_2024}. In nonlinear integrated photonics, silicon nitride (\ce{Si3N4}) has emerged as one of the dominant platforms given its wide transparency window~\cite{MunozSensors2017}, low propagation loss~\cite{BoseLSA2024}, and high refractive index~\cite{ye_low-loss_2019}. Furthermore, the maturity of the platform also enables high-yield fabrication across full \qty{300}{\mm} wafers~\cite{ou_300_2025}. The guiding and cladding materials of \ce{Si3N4} and silicon dioxide (\ce{SiO2}) exhibit distinct thermo-refractive coefficients (TRCs, $\mathrm{d}n/\mathrm{d}T$)~\cite{arbabi_measurements_2013}, while the mode confinement, and hence fraction of the optical mode that samples the two materials, depends on the optical frequency. This interplay leads to natural variation in frequency tuning across different frequencies (chromatic dispersion), which plays a crucial role in broadband nonlinear interactions spanning up to an octave or beyond, such as second-harmonic generation (SHG)~\cite{nitiss_optically_2022}, optical parametric oscillation (OPO)~\cite{sun_advancing_2024}, four-wave mixing Bragg-scattering  (FWM-BS)~\cite{li_efficient_2016}, and third-harmonic generation~\cite{carmon_visible_2007, surya_efficient_2018}. Such chromatic dispersion enables different tuning rates for the resonances associated with the input and frequency-converted output, enabling perfect energy and momentum conservation through temperature tuning~\cite{lu_efficient_2021}. Existing literature has established a strong foundation for characterizing temperature tuning in integrated photonic devices within specific frequency bands, including TRCs for \ce{Si3N4} and (or) \ce{SiO2} at individual wavelengths such as \qty{1550}{\nm}~\cite{arbabi_measurements_2013}, \qty{1510}{\nm}~\cite{zanatta_thermo_2013}, \qty{880}{\nm}~\cite{elshaari_thermo-optic_2016}, \qty{620}{\nm}~\cite{zanatta_thermo_2013}, and some have investigated their temperature dependence~\cite{johnson_determination_2022}. While these works provide essential information for specific applications, expanding toward multi-wavelength systems requires a more comprehensive understanding of the thermo-refractive behavior within the platform. For accurate design and efficient experimental operation, TRC chromatic dispersion must be further assessed, not only in the context of varying mode confinement, but also with respect to how material TRCs can exhibit a frequency-dependent response. Prior studies have demonstrated frequency dependence of TRCs in other materials such as standard optical glasses~\cite{rego_temperature_2023}, \ce{SiO2-ZrO2}~\cite{kang_wavelength_2006}, and \ce{InGaAsP}~\cite{melati_wavelength_2016}. However, broadband material TRC characterization within the \ce{Si3N4}/\ce{SiO2} integrated photonics platform has yet to be demonstrated.

  In this work, we investigate the fundamental mechanism and impact of chromatic dispersion of \ce{Si3N4}/\ce{SiO2} material TRCs across an optical octave. We provide a physical interpretation of the origin of this dispersion based on a derivative Sellmeier Model realized through temperature-dependent Lorentz oscillators. Using \ce{SiO2}-encapsulated \ce{Si3N4} microring resonators, we characterize the thermal response at five distinct laser bands where continuously tunable lasers are readily available: \qty{1550}{\nm}, \qty{1320}{\nm}, \qty{1050}{\nm}, \qty{950}{\nm}, and \qty{780}{nm}. We demonstrate that the material TRCs of both \ce{Si3N4} and \ce{SiO2} exhibit significant chromatic dispersion that must be taken into account for accurate temperature-tunable device design and modeling. To that end, we experimentally showcase the impact of TRC chromatic dispersion in the context of photoinduced SHG in Si$_3$N$_4$ microrings~\cite{lu_efficient_2021,nitiss_tunable_2023}. Our findings resolve long-standing discrepancies in frequency-matching in nonlinaear microresonators~\cite{xue_thermal_2016, afridi_effect_2023} and provide a broadly applicable workflow for high-efficiency, multi-wavelength nonlinear integrated photonics processes.

\section{Results}

\subsection*{Origin of Dispersive TRCs via Temperature-Dependent Lorentz Oscillators}

\begin{figure}[!t]
     \centering
     \includegraphics[width=\columnwidth]{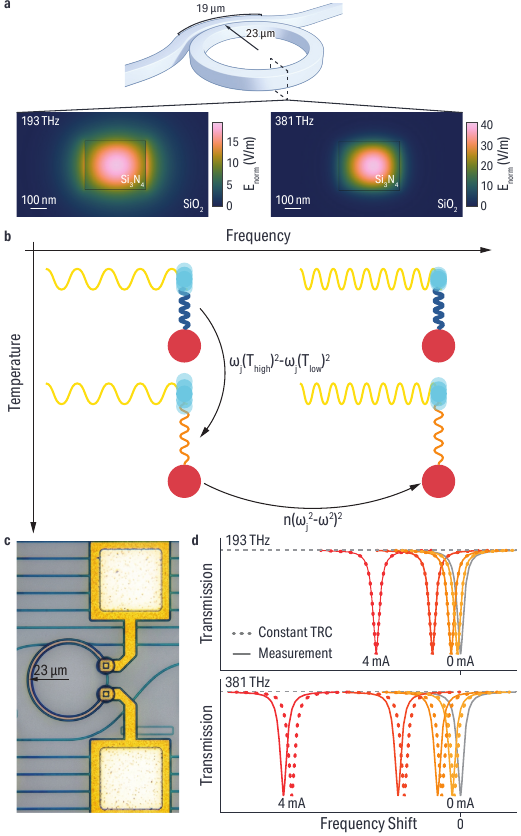}
     \caption{\label{fig:1} Physical origins and experimental observation of TRC chromatic dispersion. %
     \textbf{a} Modal confinement profiles within the \ce{Si3N4} microring resonator for low-frequency (left) and high-frequency (right) signals. The geometric contribution to the dispersive thermal response in effective refractive index arises from frequency-dependent mode overlap with the core and cladding materials. %
     \textbf{b} Schematic of the Lorentz oscillator model illustrating the atomic-scale origin of material TRC dispersion. When temperature increased, lattice vibrations reduce the effective ``spring constant'' (restoring force) of the electron-nucleus bond, red-shifting the natural absorption frequency. The refractive index response to this shift is inherently frequency-dependent and therefore the TRC should also exhibit chromatic dispersion. %
     \textbf{c} Optical microscope image of the \ce{SiO2}-encapsulated \ce{Si3N4} microring resonator featuring an integrated metallic heater embedded in the \ce{SiO2} cladding above the \ce{Si3N4} core. %
     \textbf{d} Expected measured resonance frequencies (solid curves) under varying injection currents near 193~THz (top) and 381~THz (bottom). The expected frequencies include dispersion in both the modal confinement and the material TRCs, and are compared to the case where only modal confinement dispersion is considered (dashed lines), and constant material TRCs near \qty{1550}{\nm} are assumed. At 381~THz, there is a clear additional red-shift due to dispersion in the material TRCs.} %
     
\end{figure}

  
  To begin our study of dispersive TRCs, we first consider the geometric contribution governed by the evolution of the optical mode profile across an octave-spanning range [\cref{fig:1}a]. The effective thermo-refractive coefficient of the microring resonator can be related to the material TRCs of the \ce{Si3N4} core ($\mathrm{d}n_\mathrm{Si_3N_4}/\mathrm{d}T$) and \ce{SiO2} cladding ($\mathrm{d}n_\mathrm{SiO_2}/\mathrm{d}T$) and the overlap factors of the mode with the core ($\Gamma_\mathrm{Si_3N_4}$) and cladding ($\Gamma_\mathrm{SiO_2}$), simialr to that defined in ref.~\cite{robinson_first-principle_2008, robinson_-chip_2008}, respectively~\cite{nejadriahi_thermo-optic_2020}:
  \begin{align}
    \frac{\mathrm{d}n_\mathrm{eff}}{\mathrm{d}T}(\nu) = \Gamma_\mathrm{Si_3N_4}(\nu)\frac{\mathrm{d}n_\mathrm{Si_3N_4}}{\mathrm{d}T} + \Gamma_\mathrm{SiO_2}(\nu)\frac{\mathrm{d}n_\mathrm{SiO_2}}{\mathrm{d}T}
  \end{align}
  As the optical frequency increases, the $\mathrm{TE}_\mathrm{0}$ mode experiences enhanced spatial confinement, shifting the power fraction from the \ce{SiO2} cladding into the \ce{Si3N4} core. Given the material TRC of \ce{Si3N4} is higher than that of \ce{SiO2} (e.g., \qty{\mathrm{d}n/\mathrm{d}T \approx 2.45 \times 10^{-5}}{\per\K} vs. \qty{\mathrm{d}n/\mathrm{d}T \approx 1 \times 10^{-5}}{\per\K} at \qty{193}{\THz})~\cite{arbabi_measurements_2013, moille_kerr-microresonator_2019}, this redistribution leads to a natural increase in the effective thermo-refractive coefficient~\cite{kang_high_2017}. While this geometric dispersion is well-documented in high-index-contrast platforms~\cite{deloach_thermal_2025, yang_athermal_2022}, it only accounts for a fraction of the total frequency resonance shifts observed in experiments. This necessitates a more fundamental investigation into the chromatic dispersion of the material TRCs themselves. 
  
  To understand the physical origin of dispersive material TRCs, we examine the material’s response through the Lorentz oscillator model [\cref{fig:1}b]. When light propagates through a dielectric medium, the oscillating electric field induces a light-matter interaction that can be modeled as a displacement of electrons bound to atomic nuclei via a restoring force. This driven oscillation results in a frequency- and temperature-dependent susceptibility~\cite{zhang_infrared_1998, gruner_semiconductors_2002}:
  \begin{align}
    \chi(\omega, T) = \sum_\mathrm{j} \frac{N_\mathrm{j}(T)e^\mathrm{2}}{\epsilon_\mathrm{0}m} \frac{1}{\omega_\mathrm{j}(T)^\mathrm{2}-\omega^\mathrm{2}-i\gamma_\mathrm{j}\omega}
  \end{align}
  where $N_\mathrm{j}(T)$ is the electron densities, $e$ is the elementary charge, $m$ is the electron mass, and $\omega_\mathrm{j}(T)$ represents the natural resonance (absorption) frequencies of the material. In the transparent regime, the damping coefficient $\gamma_\mathrm{j}$ is negligible, allowing the susceptibility to be simplified: 
  \begin{align}
    \chi(\omega, T) = \sum_\mathrm{j} \frac{N_\mathrm{j}e^\mathrm{2}}{\epsilon_\mathrm{0}m} \frac{1}{\omega_\mathrm{j}(T)^\mathrm{2}-\omega^\mathrm{2}}
  \end{align}
  From the susceptibility, we derive the refractive index $n(\omega,T) = \sqrt{1+\chi(\omega,T)}$. This leads directly to the Sellmeier model~\cite{sellmeier_ueber_1872}, which describes the refractive index through a summation of resonance contributions:
  \begin{align}\label{eq:Sellmeier formulation}
    n(\omega,T) =
        \sqrt{1+ \sum_\mathrm{j}{\frac{A_\mathrm{j}(T)}{s_\mathrm{j}(T)-\omega^\mathrm{2}}}} 
  \end{align}
  where $\ A_\mathrm{j}(T) \equiv N_\mathrm{j}(T)e^\mathrm{2}/\epsilon_\mathrm{0}m$ is the square of the plasma frequency and $s_\mathrm{j}(T) \equiv \omega_\mathrm{j}(T)^\mathrm{2} $ is the square of the absorption frequency. By considering the temperature dependence of these parameters, specifically the shift in the electronic absorption edge, we have a physical framework for the chromatic dispersion of the material TRC.

  As the material temperature increases, two primary physical mechanisms compete to determine the thermo-refractive response. On one hand, the electron density $N_\mathrm{j}(T)$ typically decreases due to thermal expansion of volume, which tends to lower the refractive index. On the other hand, the natural resonance frequencies $\omega_\mathrm{j}$ experience a red-shift due to the softening in the restoring force of the electron-nucleus bond and increases the electronic polarizability. In solid-state contexts, this mechanism is equivalently expressed as a temperature-dependent redshift of the electronic bandgap or excitonic transition energy, as captured in bandgap-based dispersion models such as the Ghosh Model~\cite{ghosh_model_1995, ghosh_sellmeier_1997}. For the small temperature perturbations used in our characterization, the volume expansion effect ($\partial A_\mathrm{j}/\partial T$) is negligible, allowing us to simplify the TRC expression to be represented by the impact of the absorption frequency shift:
  \begin{align}\label{eq:Sellmeier model}
    \frac{\partial{n}}{\partial{T}} =& \sum_\mathrm{j}{\frac{\partial{n}}{\partial{s_\mathrm{j}}}\frac{\partial{s_\mathrm{j}}}{\partial{T}}} + \sum_\mathrm{j}{\frac{\partial{n}}{\partial{A_\mathrm{j}}}\frac{\partial{A_\mathrm{j}}}{\partial{T}}}
    \nonumber\\
    \approx& -\frac{1}{2n} \sum_\mathrm{j}{\frac{A_\mathrm{j}}{(s_\mathrm{j}-\omega^2)^2}\frac{\partial{s_\mathrm{j}}}{\partial{T}}}
  \end{align}
  This temperature-induced red-shift in the resonance frequency ($\partial s_\mathrm{j}/\partial T < 0$) yields a positive TRC, which in turn drives the observed red-shift of the cavity resonances.

  Crucially, this expression reveals the origin of the chromatic dispersion in the material response. While the oscillator strength $A_\mathrm{j}$ and the absorption frequency shift rate $\partial s_\mathrm{j} / \partial T$ are frequency-independent constants of the material, the term $(s_\mathrm{j} - \omega^\mathrm{2})^\mathrm{2}$ in the denominator is  sensitive to the frequency of the propagating light. As the optical frequency $\omega$ approaches the material's UV absorption edge $\sqrt{s_\mathrm{j}}$~\cite{corato-zanarella_absorption_2024, taft_characterization_1971, nekrashevich_electronic_2014}, the detuning from the absorption resonance decreases, causing a non-linear acceleration in the TRC value. By incorporating this chromatic dispersive material TRC, we expect to observe an experimental deviation from predictions based on constant TRC values [\cref{fig:1}c-d]. This intrinsic material TRC dispersion enhances the geometric effects of modal confinement.

\subsection*{Broadband resonance mapping and extraction of dispersive material TRCs}

\begin{figure*}[t]
     \centering
     \includegraphics{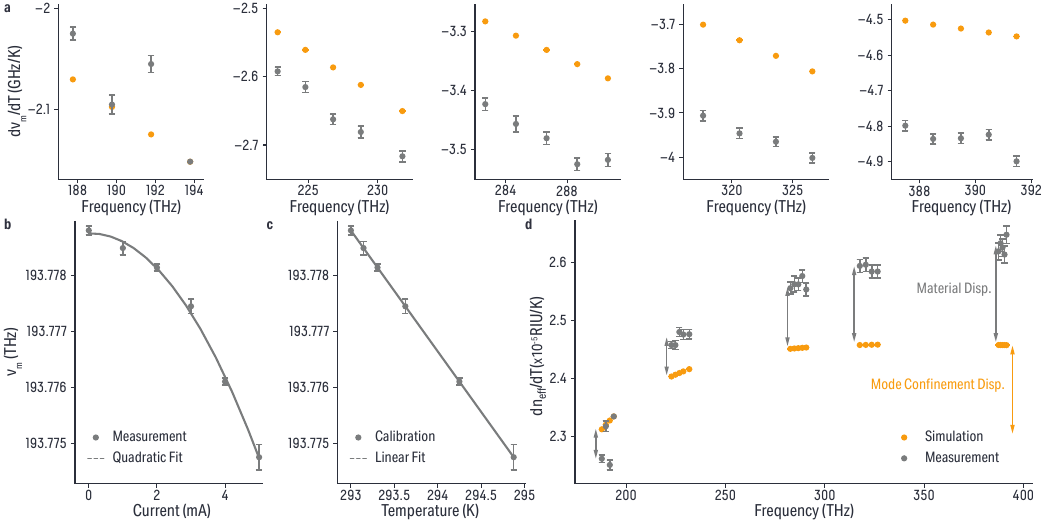}
     \caption{\label{fig:2}
     Spectroscopic characterization and calibration of resonance frequency thermal responses revealing chromatic material TRCs. %
     \textbf{a} Measured (gray) and simulated (orange) resonance frequency shifts per unit temperature across multiple laser bands (\qty{1550}{\nm}, \qty{1320}{\nm}, \qty{1050}{\nm}, \qty{950}{\nm}, and \qty{780}{\nm}), where the simulations assume constant material TRCs. While  measurement aligns with simulation around \qty{193}{\THz}, a significant discrepancy emerges at higher frequencies. %
     \textbf{b} Experimentally measured resonance frequency as a function of injection current ($I$), showing quadratic dependence consistent with Joule heating, where the temperature change follows:  $\Delta T \propto P = I^\mathrm{2} R$~\cite{bahadori_thermal_2018} with $P$ representing the power, $I$ representing the current, and $R$ representing the resistance. %
     \textbf{c} Calibration of measured resonance frequencies against FEM-simulated frequency versus temperature curves near \qty{193}{\THz} to map injection current to absolute local temperature. %
     \textbf{d} Measured (gray) and simulated (orange) change in effective refractive index per unit temperature across an octave. The experimental result reveals the additional level of dispersion beyond the modal confinement effects captured in the simulation following constant material TRCs, originating from the intrinsic chromatic dispersion of the material TRCs. The uncertainty bars in (a), (d) represent one standard deviation values (68~\% confidence interval) from data fitting of the slope of resonance frequency versus temperature. The uncertainty bars in (b)-(c) represent the combined standard uncertainty considering both the one standard deviation values (68~\% confidence intervals) of the measurement samples and the manufacturer uncertainty of \qty{\pm 60}{\MHz} from the wavemeter resolution. %
     }
\end{figure*}

\begin{figure}[t]
     \centering
     \includegraphics[width=\columnwidth]{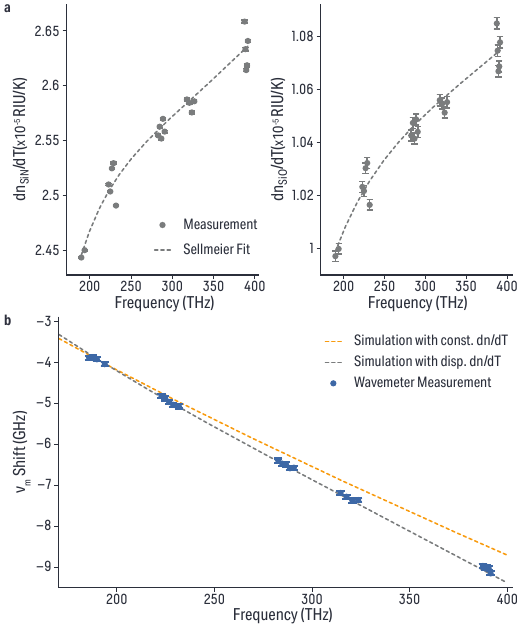}
     \caption{\label{fig:3} Extraction of octave-spanning chromatic material TRCs and experimental validation of the dispersive TRC model. %
     \textbf{a} Extracted TRCs for \ce{Si3N4} (left) and \ce{SiO2} (right) across an octave. Both materials exhibit clear chromatic behavior with a \qty{\approx 7}{\percent} variation between \qty{193}{\THz} and \qty{391}{\THz}. Equations derived from the derivative Sellmeier model are fitted against the experimental results to extract the coefficients required for accurate modeling of resonance thermal response with dispersive TRCs. %
     \textbf{b} Comparison of measured resonance frequency shifts ($\Delta I =$ \qty{5}{\mA}, $\Delta T \approx$ \qty{1.87}{\K}) across an octave (blue dots) against FEM simulations. The experimental data shows excellent agreement with the dispersive TRC model (gray dashed), whereas the constant TRC model (orange dashed) fails to capture the trend. All reported uncertainties correspond to one standard deviation. %
     }
\end{figure}
    
  To investigate the TRCs of \ce{Si3N4} and \ce{SiO2}, we employ \ce{SiO2}-encapsulated \ce{Si3N4} microring resonators with a thickness of \qty{H=670}{\nm} and a ring radius of \qty{RR=23}{\um}, featuring ring widths of \qty{RW=840}{\nm} and \qty{RW=830}{\nm}. The integrated heaters embedded within the \ce{SiO2} cladding above the \ce{Si3N4} allow precise local temperature control of the system via electrical current injection [\cref{fig:1}c]. %
  Critically, these resonators are coupled to pulley-like bus waveguides, which ensure efficient coupling to the fundamental transverse-electric mode ($\mathrm{TE}_\mathrm{0}$) of the cavity across an octave-spanning range from \qty{185}{\THz} to \qty{391}{\THz}~\cite{moille_broadband_2019}. %
  
  We characterize the frequency shift of the $\mathrm{TE}_\mathrm{0}$ resonances by sweeping the injection current from \qty{0}{\mA} to \qty{5}{\mA}. This current range induces measurable resonance frequency shifts within a small temperature window. %
  While the heaters are capable of reaching temperatures exceeding \qty{520}{\K}, we restrict the initial characterization to low injection currents to avoid the higher-order temperature-dependent nonlinear thermo-refractive response reported in previous studies~\cite{johnson_determination_2022}. %
  
  The frequency shifts per unit temperature, $\mathrm{d}\nu_\mathrm{m,meas}/\mathrm{d}T$, were measured across the five wavelength bands and compared with simulated values, $\mathrm{d}\nu_\mathrm{m,sim}/\mathrm{d}T$ assuming constant TRCs [\cref{fig:2}a]. To extract the $\mathrm{d}\nu_\mathrm{m,meas}/\mathrm{d}T$ values, we calibrate the local temperature for a given injection current and employ a reference-mapping technique. Here, we assume a uniform local temperature across the optical mode, an approximation validated by thermal finite element method (FEM) simulations provided in Supplementary Information \cref{supsec:localtemp}. We compare the experimentally measured resonance frequencies [\cref{fig:2}b] with simulated values obtained from FEM simulations around \qty{193}{\THz}, where the TRCs of \ce{Si3N4} and \ce{SiO2} are well-established~\cite{arbabi_measurements_2013}. %
  By utilizing the standard constant TRCs over a small temperature range (\qty{293.15}{\K} to \qty{298.15}{\K}), we determine the theoretical $\mathrm{d}\nu/\mathrm{d}T$ from the slope of the resonance frequency tuning with temperature near \qty{193}{\THz}. %
  The experimental frequency shifts near \qty{193}{\THz} are then mapped to this reference slope to establish a precise current-to-temperature calibration. Using this calibration, we are able to extract $\mathrm{d}\nu_\mathrm{m,meas}/\mathrm{d}T$ across the remaining wavelength bands [\cref{fig:2}c]. While the measurement aligns well with simulation at the \qty{193}{\THz} reference point, a substantial discrepancy emerges and grows as the resonance frequency increases, aligning with our expectation from the prior section. This widening offset suggests that the standard assumption of constant material TRCs for \ce{Si3N4} and \ce{SiO2} is insufficient for broadband modeling, signaling the presence of intrinsic chromatic dispersion in the material TRCs. 

  To quantify the significance of this observation, we further extract the effective thermo-refractive coefficient of the microring resonator, $\mathrm{d}n_\mathrm{eff}/\mathrm{d}T$. The relationship between the resonance frequency shift and the effective index change is given by~\cite{qiu_athermal_2015, weituschat_photonic_2020}:
  \begin{align}
    \frac{\mathrm{d}n_\mathrm{eff}(\nu_\mathrm{m},T)}{\mathrm{d}T}=-\frac{n_\mathrm{g}(\nu_\mathrm{m})}{\nu_\mathrm{m}(T)}\frac{\mathrm{d}\nu_\mathrm{m}(T)}{\mathrm{d}T}-n_\mathrm{eff}(\nu_\mathrm{m})\alpha(T)
  \end{align}
  where $n_\mathrm{g}(\nu_\mathrm{m})$ is the group index at the resonance frequency $\nu_\mathrm{m}$, $\nu_\mathrm{m}(T)$ is the temperature-dependent resonance frequency, and $\alpha(T)$ is the thermal expansion coefficient of the microring resonator. Since our measurements are conducted over a narrow temperature range, we neglect the thermal expansion term and the temperature dependence of $\nu_\mathrm{m}$ in the denominator of the first term in the expression, resulting in: 
  \begin{align}
    \frac{\mathrm{d}n_\mathrm{eff}(\nu_\mathrm{m})}{\mathrm{d}T}\approx-\frac{n_\mathrm{g}(\nu_\mathrm{m})}{\nu_\mathrm{m}}\frac{\mathrm{d}\nu_\mathrm{m}}{\mathrm{d}T}
  \end{align}
  By applying this relation to both our experimental and simulated data, we reveal the two distinct contributions to TRC dispersion within the system [\cref{fig:2}d]. The simulated curve inherently exhibits a dispersive behavior of increasing $\mathrm{d}n_\mathrm{eff}/\mathrm{d}T$ at higher frequencies. This trend originates solely from geometric mode confinement~\cite{wang_mode_2023, pruiti_thermo-optic_2020, deloach_thermal_2025}. However, the appreciable deviation between the experimental measurements and this purely geometric dispersion effect, which grows as a function of frequency, confirms that beyond the well-understood modal confinement effects, there exists a second contribution due to dispersion of the material TRCs. Notably, the variation of $\mathrm{d}n_\mathrm{eff}/\mathrm{d}T$ originated from material TRCs is 1.3 times that from the modal confinement.
  
  To isolate the individual material TRCs from the composite resonance shift, we follow a similar formulation presented in~\cite{arbabi_measurements_2013}, where the total thermal response can be expressed as a linear combination of the constituent material TRCs:
  \begin{align}
    \frac{\mathrm{d}\nu_\mathrm{RW1}}{\mathrm{d}T} = \frac{\mathrm{\partial}\nu_\mathrm{RW1}}{\mathrm{\partial}n_\mathrm{core}}\frac{\mathrm{d}n_\mathrm{core}}{\mathrm{d}T} + \frac{\mathrm{\partial}\nu_\mathrm{RW1}}{\mathrm{\partial}n_\mathrm{clad}}\frac{\mathrm{d}n_\mathrm{clad}}{\mathrm{d}T},\\
    \frac{\mathrm{d}\nu_\mathrm{RW2}}{\mathrm{d}T} = \frac{\mathrm{\partial}\nu_\mathrm{RW2}}{\mathrm{\partial}n_\mathrm{core}}\frac{\mathrm{d}n_\mathrm{core}}{\mathrm{d}T} + \frac{\mathrm{\partial}\nu_\mathrm{RW2}}{\mathrm{\partial}n_\mathrm{clad}}\frac{\mathrm{d}n_\mathrm{clad}}{\mathrm{d}T}.
  \end{align}
  where the sensitivities, $\mathrm{d}\nu/\mathrm{d}n$, represent the dependence of the resonance frequency on the core and cladding refractive indices, respectively, derived from FEM simulations. While previous studies have resolved these coefficients by comparing fundamental transverse electric ($\mathrm{TE}$) and transverse magnetic ($\mathrm{TM}$) modes, our microrings do not support the $\mathrm{TM}_\mathrm{0}$ mode across the entire octave-spanning range. Consequently, we solve the system of equations by measuring the resonance frequency shifts of two adjacent microrings with distinct ring widths ($\mathrm{RW1}$ and $\mathrm{RW2}$) on the same chip. This approach enables measurements across the octave using resonators that possess identical thickness, material profile, and thermal environments, while providing the linearly independent sensitivities required to extract TRCs.
  
  The extracted material TRCs are presented in [\cref{fig:3}a]. The gray markers represent the experimentally extracted TRCs derived from the measured resonance frequency shifts and simulated sensitivities. We observe a distinct and consistent increase in the TRCs for both \ce{Si3N4} and \ce{SiO2} as the frequency increases. Notably, the TRCs at \qty{391}{\THz} are approximately \qty{7}{\percent} larger than those at the \qty{193}{\THz} reference. This quantified variation provides definitive evidence of intrinsic chromatic dispersion in the material thermo-refractive response, independent of waveguide geometry. %

  To validate our theoretical framework, we fit the derivative Sellmeier model \cref{eq:Sellmeier model} to the experimentally extracted TRCs [\cref{fig:3}a]:
  \begin{align}\label{eq:Final Sellmeier model}
     \frac{d{n}}{d{T}} = -\frac{1}{2n} \sum_\mathrm{j}{\frac{A_\mathrm{j}}{(s_\mathrm{j}-\omega^2)^2}\frac{d{s_\mathrm{j}}}{d{T}}}
  \end{align}
  where the refractive indices $n$ follow the established Sellmeier model parameters for \ce{Si3N4}~\cite{luke_broadband_2015} and \ce{SiO2}~\cite{baak_silicon_1982}. We find that the model accurately captures the dispersive behavior of the TRCs across the entire octave, for which the fitted coefficients can be found in Supplementary Information \cref{supsec:coefficients}. We subsequently implemented these dispersive TRCs into FEM simulation to compare the predicted resonance shifts against the standard constant TRC simulation. The evaluation was performed by calculating the shift between room temperature (\qty{293.15}{\K}) and the calibrated temperature corresponding to a \qty{5}{\mA} injection current (\qty{295.02}{\K}). As shown in [\cref{fig:3}b], the experimental frequency shifts across all five wavelength bands align with the dispersive model. In contrast, the constant TRC simulation exhibits a significant and growing divergence at higher frequencies.

  Notably, the impact of the material dispersion is significant, where a temperature change of less than \qty{2}{\K} near \qty{391}{\THz} results in a frequency shift discrepancy of approximately \qty{500}{\MHz}. This error is more than 2 times the resonance linewidth of the microring resonator, revealing that typical constant TRC model is insufficient for applications requiring high-precision frequency control and matching. These results highlight the necessity of incorporating dispersive TRCs in the design of temperature-tunable integrated photonic devices, particularly for multi-wavelength nonlinear processes. %

\subsection*{Impact of the Dispersive TRC Model on Second-Harmonic Generation}

\begin{figure*}[t]
     \centering
     \includegraphics{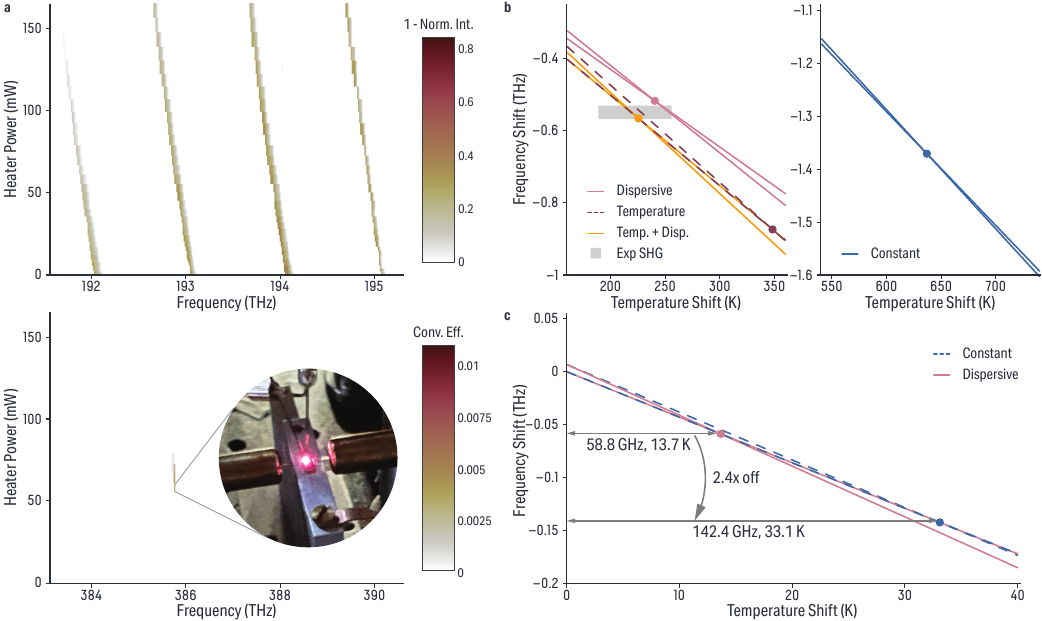}
     \caption{\label{fig:4} Second harmonic generation and TRC model validation. %
     \textbf{a} Experimental search for SHG in a \ce{Si3N4} microring resonator with \qty{\approx 700}{\mW} on-chip optical power near \qty{193}{\THz}. Transmission heat map in the fundamental band (top) and generated light in the second harmonic band (bottom), taken as a function of frequency (x-axis) and heater power (y-axis). Here, the plotted FH power is normalized against the maximum transmission power of the \qty{193}{\THz} signal and the SHG conversion efficiency is normalized to the on-chip power of the \qty{193}{\THz} pump, revealing an SHG signal at \qty{192.89}{\THz} with integrated heater power of \qty{57.9}{mW}. Bottom inset: Photograph showing visible red-light scattering from the ring during SHG. %
     \textbf{b} FEM simulation of the microring resonator that demonstrated SHG while considering the resonance frequency matching condition under four scenarios: constant (blue solid), dispersive (pink solid), temperature-dependent (red dashed), and combined temperature-dependent and dispersive TRCs (orange solid). In each scenario, a pair of tuning curves is shown, where the shallower slope represents the FH resonance, while the steeper slope corresponds to the SH resonance divided by two to map to the FH regime. The intersection points represent the predicted temperature and frequency shifts required to align FH and SH resonances. The combined model aligns best with experimental data, while the constant TRC model exhibits a significant temperature offset exceeding 400 K. %
     \textbf{c} Sensitivity analysis of a microring resonator design with a small initial frequency mismatch. Even in this low-temperature regime, the constant TRC model overestimates the required temperature and frequency shift by a factor of 2.4 compared to the dispersive model.  %
     }
\end{figure*}

  The chromatic dispersion of material TRCs is very consequential in multi-band nonlinear processes such as SHG, OPO, and FWM-BS. In a \ce{Si3N4} ring resonator, achieving efficient SHG requires frequency matching between the fundamental (FH) and second-harmonic (SH) modes, where the latter is at twice the frequency of the former. Additionally, the frequency-matched modes need to be phase-matched, though all-optical poling mediated by the coherent photogalvanic process in \ce{Si3N4} enables automatic quasi-phase matching and relaxes this condition~\cite{nitiss_optically_2022}. While high quality factor resonators with small mode volumes (large free spectral ranges) can realize high normalized and absolute conversion efficiencies~\cite{lu_efficient_2021}, fabrication tolerances and the relatively scarce mode density mean that temperature tuning is typically needed to achieve frequency matching. This is possible due to the difference in tuning rates of the FH and SH resonances with temperature, due to TRC dispersion. Thus, accurate prediction of the frequency-matching temperature and resonance frequency serves as a test for the dispersive material TRC model presented in this work.

  We experimentally characterized the SHG performance of our \ce{Si3N4} microring resonator by sweeping the pump frequency and local temperature via the integrated heater [\cref{fig:4}a], while monitoring the generation of light in the SH band~[\cref{fig:4}b]. As the heater power increased, we tracked four distinct resonance modes. At a heater power of \qty{57.9}{\mW}, we observed a strong SHG signal with the pump centered at $\approx$\qty{192.89}{\THz}. This was accompanied by a sharp increase in intensity in the SH band and visible red-light scattering from the microring circumference (see inset, [\cref{fig:4}b). This experimental SHG observation serves as a critical benchmark for our theoretical framework. By comparing the frequency shift required to achieve resonance alignment against our predictive model, we validate the essential role of material TRC dispersion in multi-wavelength integrated photonics.

  To demonstrate the necessity of dispersive material TRCs and the accuracy of our extracted model, we performed FEM simulations. Because the device is subjected to high optical pump powers (\qty{\approx 700}{\mW}) and significant heater-induced tuning, we take into account the nonlinear temperature dependence of the TRCs. Following prior study~\cite{johnson_determination_2022}, we incorporated up to the second-order temperature terms for both \ce{Si3N4} and \ce{SiO2}: 
  \begin{align}
    \frac{\partial{n_\mathrm{Si_3N_4}}}{\partial{T}} =& (2.14\cdot10^\mathrm{-11}K^\mathrm{-3} \cdot T^\mathrm{2}+2.02\cdot10^\mathrm{-8}K^\mathrm{-2} \cdot T \nonumber\\ 
    &+1.71\cdot10^\mathrm{-5}K^\mathrm{-1})\\
    \frac{\partial{n_\mathrm{SiO_2}}}{\partial{T}} =& (1.89\cdot10^\mathrm{-11}K^\mathrm{-3} \cdot T^2+1.07\cdot10^\mathrm{-8}K^\mathrm{-2} \cdot T \nonumber\\
    &+7.48\cdot10^\mathrm{-7}K^\mathrm{-1})
  \end{align}
  We assume a universal temperature scaling across the octave, normalizing the nonlinear increase in TRCs to our extracted dispersive values. This approach allows us to evaluate four distinct modeling scenarios: (i) constant TRC, (ii) dispersive TRC only, (iii) temperature-dependent TRC only, and (iv) the comprehensive TRC model incorporating both dispersive and temperature-dependent terms. 

  The simulated frequency shifts for the FH and SH resonances are shown in [\cref{fig:4}b], with the zero-shift reference set to the intrinsic FH resonance frequency while the SH resonance frequency is divided by two to map to the FH domain. The intersection of these two temperature tuning curves represents the predicted temperature and frequency shift required to achieve the spectral overlap necessary for efficient SHG. Our analysis reveals that incorporating the dispersive material TRCs is essential for achieving estimates that align with experimental data. In particular, both the pure dispersive model and the comprehensive model (dispersive with temperature dependence) predict resonance frequency and temperature shifts that match the experimentally observed SHG location of \qty{549.48}{\GHz} \qty{\pm~18.84}{\GHz} resonance frequency shift and \qty{222.5}{\K} \qty{\pm~33.5}{\K} temperature shift (see Supplementary Information \cref{supsec:SHG temp} section for the estimation of the experimental SHG location and range). The pure dispersive model predicts \qty{517.81}{\GHz} and \qty{240.61}{\K}, frequency and temperature shift, respectively, while the comprehensive model predicts values of \qty{566.34}{\GHz} and \qty{225.66}{\K}, respectively, which match the experiments to within their uncertainties. 
  
  In contrast, models that neglect material TRC chromatic dispersion fail to provide a physically plausible result. The constant-TRC model predicts a significantly larger frequency shift of \qty{1370.72}{\GHz}, resulting in a required temperature shift of \qty{636.92}{\K}, which is about \qty{400}{\K} higher than our two dispersive models, and far outside the range of our experimental observations. Even taking into account the temperature-dependent corrections, the frequency shift is \qty{873.87}{\GHz}, with a temperature shift over \qty{110}{\K} larger than the dispersive models. These results suggest that while nonlinear temperature dependence refines the thermal budget estimation, it cannot compensate for the absence of chromatic material dispersion. The frequency-matching conditions for broadband nonlinear processes can only be accurately predicted by implementing the dispersive TRC model established in this work.

  To further isolate the influence of chromatic dispersion from high-temperature nonlinearities, we modeled a \ce{Si3N4} microring resonator specifically designed with a minimal intrinsic frequency mismatch for SHG. By reducing the required tuning range, we ensure the device operates in a regime where the temperature dependence of the TRCs is negligible, thereby providing a clear comparison between the constant and dispersive frameworks. As illustrated in [\cref{fig:4}c], we implemented the constant and dispersive TRCs to evaluate the tuning required to overcome an intrinsic mismatch of only \qty{3.4}{\GHz}. Our dispersive TRC model predicts that a FH frequency shift of \qty{29.4}{\GHz} at a temperature increase of \qty{13.7}{\K} is sufficient to achieve frequency matching for SHG. In contrast, the standard constant-TRC model predicts a frequency shift and temperature requirement of \qty{71.2}{\GHz} and \qty{33.1}{\K}, respectively, a factor 2.4$\times$ higher for each quantity. These results further reinforce the importance of using the dispersive material TRCs we have unveiled for optimization of wideband nonlinear integrated photonic devices.
  

\section{Discussion}

  In this work, we study the physical origin and implications of chromatic material TRCs in \ce{Si3N4}/\ce{SiO2} photonics. We have demonstrated that they exhibit a clear chromatic dispersion across an optical octave, with a variation of approximately \qty{7}{\percent} between \qty{185}{\THz} and \qty{391}{\THz}. While traditional design workflows accurately account for geometric mode confinement dispersion, our findings reveal that the omission of material TRC dispersion leads to a significant discrepancy in predicting multi-wavelength nonlinear interactions, exemplified by a phase-matching temperature discrepancy of over \qty{400}{\K} for SHG. While we have focused on the \ce{Si3N4}/\ce{SiO2} platform, our methodology should apply to a host of other nonlinear integrated photonics platforms~\cite{dutt_nonlinear_2024}, such as Aluminum Nitride and Lithium Niobate, whose electronic absorption edges are also in the ultraviolet or visible regimes.

  To that end, our findings carry broad implications for nonlinear photonics. Beyond the SHG demonstration in this work, the efficiency of processes such as OPO and FWM-BS depends critically on precise resonance alignment across multiple spectral bands. Furthermore, in applications like metrology using octave-spanning optical frequency combs, accurate temperature control is essential for phase-locking comb teeth to fixed atomic references or stabilizing the carrier-envelope offset frequency via $f-2f$ self-referencing. Our analysis reveals that standard constant material TRC models incorrectly estimate these thermal frequency shifts by more than a factor of two, even in low-temperature regimes where nonlinear thermal effects are negligible. By incorporating the dispersive TRC framework, researchers can significantly reduce the reliance on extensive post-fabrication trial-and-error tuning. This enables a more predictable workflow, ensuring that the thermal budget and heater architectures are correctly designed to achieve a target frequency, leading to more robust thermal control for multi-wavelength integrated photonic circuits.

\section*{Acknowledgments}
\noindent S-C.O., K.S., and G.M. acknowledge partial funding support from the Space Vehicles Directorate of the Air Force Research Laboratory and the NIST-on-a-chip program of the National Institute of Standards and Technology.

\section*{Author Declarations}

\subsection*{Conflict of Interest}
The authors have no conflicts to disclose.


\subsection*{Data Availability}
The data that support the plots within this paper and other findings of this study are available from the corresponding author upon reasonable request.


%

\captionsetup[figure]{name=Extended Data Fig}
\setcounter{figure}{0}
\onecolumngrid

\clearpage

\renewcommand{\appendixpagename}{\Large\centering Supplementary Information: \mytitle\vspace{0em}}
\appendix  
\appendixpage

\captionsetup[figure]{name=Fig.}
\setcounter{figure}{0}
\setcounter{equation}{0}
\renewcommand{\thesection}{S.\arabic{section}}
\renewcommand\thefigure{S.\arabic{figure}}    
\numberwithin{equation}{section}
\renewcommand\theequation{S.\arabic{equation}}


\enabletocentries

\section{Microring resonator design}

  The photonic chips were fabricated following the process presented in ref.~\textcite{MoilleNature2023} in a commercially available foundry. For the broadband resonance spectroscopy, we utilized \ce{Si3N4} microring resonators embedded in \ce{SiO2} with an outer ring radius \qty{R=23}{\um} and thickness of \qty{650}{\nm}. Two distinct ring widths, \qty{RW=840}{\nm} and \qty{RW=830}{\nm} were employed to facilitate the dual-width characterization needed to isolate core and cladding TRCs. A bus waveguide (width \qty{W_\mathrm{wg}=460}{\nm}) wrapped around the ring in a pulley-like fashion~\cite{moille_broadband_2019} with coupling length \qty{L_\mathrm{c}=19}{\um} and gap \qty{G=500}{\nm} enables efficient coupling from \qty{185}{\THz} to \qty{391}{\THz}. For the SHG demonstration, we utilized a resonator with \qty{R=22.85}{\um}, thickness \qty{650}{\nm} and ring width \qty{RW=840}{\nm}. A slightly adjusted pulley waveguide with coupling length \qty{L_\mathrm{c}=17}{\um} and gap \qty{G=550}{\nm} is implemented for the device. To access SHG, approximately \qty{700}{\mW} of on-chip optical power in the fundamental band was applied. Importantly, all devices were sourced from the same wafer to maintain consistent material stoichiometry and TRC profiles across the study. 

\section{Resonance frequency versus temperature
measurement}

  The $\mathrm{TE_0}$ resonance frequencies were characterized across five distinct frequency bands using a series of continuously tunable lasers. To ensure statistical robustness and capture the local dispersion profile, five resonance modes were measured within each band. For each mode, the temperature was swept across six points, controlled by varying the integrated heater current from \qty{0}{\mA} to \qty{5}{\mA}. The precise resonance frequency for each temperature point was captured using a wavemeter (with accuracy of \qty{\pm 60}{\MHz}). Data points were recorded when the laser frequency reached the center of the resonance mode, identified by the transmission signal reaching its minimum intensity for singlet modes or the local maximum for doublet modes.

\section{Local temperature of microring resonator optical mode with integrated heater
\label{supsec:localtemp}}

\begin{figure*}[h]
    \centering
    \includegraphics{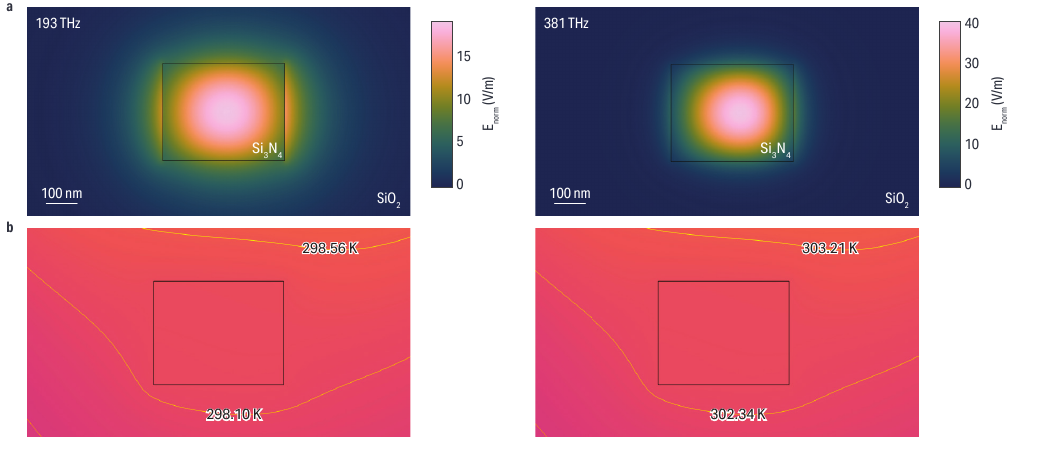}
    \caption{\label{supfig:temperature} 
    Modal confinement and thermal uniformity in the \ce{Si3N4} microring resonator. %
    \textbf{a} Simulated optical mode profiles within the \ce{Si3N4} microring resonator at low-frequency (left) and high-frequency (right) portions of the spectrum.  %
    \textbf{b} Cross-section temperature distribution within the microring resonator due to the integrated heater at the maximum spectroscopic measurement temperature (\qty{298.15}{\K}) and in a higher temperature scenario (\qty{303}{\K}). The simulations demonstrate a spatially homogeneous thermal profile across the modal volume, where temperature gradients are negligible. This spatial uniformity validates the use of a single temperature value for the thermo-refractive analysis and the extraction of material TRCs is reasonable. %
    }
\end{figure*}

To simplify the extraction of TRCs, we assume a uniform temperature distribution across the microring resonator's optical modes (\cref{supfig:temperature}a). This assumption is validated through finite-element method (FEM) simulation of the device's thermal profile with the integrated heater. At the highest temperature where we conducted the spectroscopic measurements (\qty{298.15}{\K}), the temperature variation across the modal volume is found to be negligible (\cref{supfig:temperature}b). Furthermore, simulation conducted at higher temperature (\qty{303}{\K}) exhibits a similar flat temperature profile, confirming that the mode experiences a spatially homogeneous thermal shift. Therefore, the estimation of a single temperature value for the entire modal confinement is physically sound for the thermo-refractive analysis performed in this work.

\section{Numerical extraction of resonance frequency sensitivities}

  To accurately map the shift in resonance frequency to the individual material TRCs, we numerically determine the sensitivity coefficients, $\mathrm{d}\nu/\mathrm{d}n_\mathrm{core}$ and $\mathrm{d}\nu/\mathrm{d}n_\mathrm{clad}$. These sensitivities were obtained using a FEM eigensolver. For each simulated mode and geometry, we performed a variational analysis by offsetting the refractive index of the \ce{Si3N4} core and \ce{SiO2} cladding. A total of five data points for each refractive index offset was simulated to ensure a linear response. The resulting resonance frequency shifts were then fitted to a linear model, where the slope represents the specific sensitivity of the resonance to the refractive index of the respective material.

\section{Derivative Sellmeier Model Coefficients
\label{supsec:coefficients}}

To derive the analytical expression for the dispersive thermo-refractive coefficients (TRCs), we utilize established Sellmeier parameters for \ce{Si3N4} and \ce{SiO2} from prior literature~\cite{luke_broadband_2015, baak_silicon_1982}. In the case of \ce{Si3N4}:
\begin{align}
    n_\mathrm{SiN} = \sqrt{1+\frac{3.0249\lambda^\mathrm{2}}{\lambda^\mathrm{2}-135.3406^\mathrm{2}}+\frac{40314\lambda^\mathrm{2}}{\lambda^\mathrm{2}-1239842^\mathrm{2}}}
\end{align}
The standard wavelength-dependent Sellmeier equation is first transformed into a frequency-domain Lorentz oscillator expression \cref{eq:Sellmeier formulation}:
\begin{align}
    n_\mathrm{SiN} = \sqrt{1+ \sum_j{\frac{A_\mathrm{j}}{s_\mathrm{j}-\omega^\mathrm{2}}}} =  \sqrt{1+\frac{\frac{3.0249}{135.3406^\mathrm{2}}({2\pi c})^\mathrm{2}}{\frac{({2\pi c})^\mathrm{2}}{135.3406^2}-\omega^\mathrm{2}}+\frac{\frac{40314}{1239842^\mathrm{2}}({2\pi c})^\mathrm{2}}{\frac{({2\pi c})^\mathrm{2}}{1239842^\mathrm{2}}-\omega^\mathrm{2}}}
\end{align}
This conversion provides the values for the square of the plasma frequency ($A_\mathrm{j}$) and square of the absorption frequency ($s_\mathrm{j}$) in the derivative Sellmeier model equation \cref{eq:Final Sellmeier model}: 
\begin{align}
     \frac{d{n}}{d{T}} = -\frac{1}{2n} \sum_j{\frac{A_\mathrm{j}}{(s_\mathrm{j}-\omega^2)^2}\frac{d{s_\mathrm{j}}}{d{T}}}
\end{align}
This effectively isolates the temperature-induced resonance shift ($d{s_\mathrm{j}}/{d{T}}$) as the free parameter for the experimental fit. All fixed parameters sourced from literature, alongside the resulting fitted coefficients and one standard deviation uncertainties from the data fitting are summarized in the following Table~\ref{tab:1}:
\begin{table}[h]
  \caption{
    \label{tab:1}
    Coefficients for characterizing the chromatic material TRCs of \ce{Si3N4}.
  }
  \centering
  \renewcommand{\arraystretch}{1.8}
  \begin{tabular}{l@{\hskip 0.3cm}c@{\hskip 0.3cm}c@{\hskip 0.8cm}c}
    \hline
    \ce{Si3N4 Parameter} & Coefficient & One Std. Dev. Fit Uncertainty\\
    \hline
    $A_\mathrm{1}$ &  \qty{\frac{3.0249}{(135.3406\times10^\mathrm{-9})^\mathrm{2}}({2\pi c})^\mathrm{2}}{\Hz^2}  \\
    $s_\mathrm{1}$ & \qty{\frac{({2\pi c})^\mathrm{2}}{(135.3406\times10^\mathrm{-9})^\mathrm{2}}}{\Hz^2}  \\   
    ${ds_\mathrm{1}}/{d{T}}$ & \qty{-6.44\times10^\mathrm{27}}{\Hz^2\per\K} & \qty{\pm 1.09\times10^\mathrm{25}}{\Hz^2\per\K}\\
    $A_\mathrm{2}$ & \qty{\frac{40314}{(1239842\times10^\mathrm{-9})^\mathrm{2}}({2\pi c})^\mathrm{2}}{\Hz^2} \\
    $s_\mathrm{2}$ & \qty{\frac{({2\pi c})^\mathrm{2}}{(1239842\times10^\mathrm{-9})^\mathrm{2}}}{\Hz^2} \\   
    ${ds_\mathrm{2}}/{d{T}}$ & \qty{9.57\times10^\mathrm{25}}{\Hz^2\per\K} & \qty{\pm 9.83\times10^\mathrm{24}}{\Hz^2\per\K}\\
    \hline
  \end{tabular}
\end{table}

Physically, a temperature-induced red-shift ($ds_\mathrm{j}/dT < 0$) is expected for all material resonances. However, in our derivative Sellmeier model extraction, a negative coefficient is only yielded for the resonance term closest to the experimentally surveyed frequencies (\qty{185}{\THz} to \qty{391}{\THz}), which is the first oscillator term ($j=1$) within \ce{Si3N4}. This result is due to the local dominance of the nearest absorption pole, which dictates the curvature and slope of the thermo-refractive response across the measured octave. Given the finite bandwidth measured in the experiments, the sensitivity of the fit to distant resonances is significantly reduced. Consequently, while the closest resonance term ($j=1$) captures the primary physical mechanism of the red-shift of the absorption frequency, the remaining oscillator term ($j=2$) 
incorporates the contributions of more distant resonance tails and the small, competing effects of volume expansion ($\partial A_\mathrm{j}/\partial T$) to minimize numerical residuals across the fit.

In the case of \ce{SiO2}, we follow the same derivation starting from the wavelength-dependent Sellmeier equation:
\begin{align}
    n_\mathrm{SiO2} = \sqrt{1+\frac{0.6961663\lambda^\mathrm{2}}{\lambda^\mathrm{2}-68.4043^\mathrm{2}}+\frac{0.4079426\lambda^\mathrm{2}}{\lambda^\mathrm{2}-116.2414^\mathrm{2}}+\frac{0.8974794\lambda^\mathrm{2}}{\lambda^\mathrm{2}-9896.161^\mathrm{2}}}
\end{align}
and then transform it into a frequency-domain Lorentz oscillator expression\cref{eq:Sellmeier formulation}:
\begin{align}
    n_\mathrm{SiO2} = \sqrt{1+ \sum_j{\frac{A_\mathrm{j}}{s_\mathrm{j}-\omega^\mathrm{2}}}} =  \sqrt{1+\frac{\frac{0.6961663}{68.4043^\mathrm{2}}({2\pi c})^\mathrm{2}}{\frac{({2\pi c})^\mathrm{2}}{68.4043^2}-\omega^\mathrm{2}}+\frac{\frac{0.4079426}{116.2414^\mathrm{2}}({2\pi c})^\mathrm{2}}{\frac{({2\pi c})^\mathrm{2}}{116.2414^\mathrm{2}}-\omega^\mathrm{2}}+\frac{\frac{0.8974794}{9896.161^\mathrm{2}}({2\pi c})^\mathrm{2}}{\frac{({2\pi c})^\mathrm{2}}{9896.161^\mathrm{2}}-\omega^\mathrm{2}}}
\end{align}
Similar to the case of \ce{Si3N4}, the above conversion provides the values for the square of the plasma frequency ($A_\mathrm{j}$) and square of the absorption frequency ($s_\mathrm{j}$) in the derivative Sellmeier model equation \cref{eq:Final Sellmeier model}, and isolates the temperature-induced resonance shift ($d{s_\mathrm{j}}/{d{T}}$) as the free parameter for the experimental fit.  All fixed parameters sourced from literature, alongside the resulting fitted coefficients and one standard deviation
from the data fitting are summarized in the following Table~\ref{tab:2}:
\begin{table}[h]
  \caption{
    \label{tab:2}
    Coefficients for characterizing the chromatic material TRCs of \ce{SiO2}.
  }
  \centering
  \renewcommand{\arraystretch}{1.8}
  \begin{tabular}{l@{\hskip 0.3cm}c@{\hskip 0.3cm}c@{\hskip 0.8cm}c}
    \hline
    \ce{SiO2 Parameter} & Coefficients & One Std. Dev. Fit Uncertainty \\
    \hline
    $A_\mathrm{1}$ & \qty{\frac{0.6961663}{(68.4043\times10^\mathrm{-9})^\mathrm{2}}({2\pi c})^\mathrm{2}}{\Hz^2}  \\
    $s_\mathrm{1}$ & \qty{\frac{({2\pi c})^\mathrm{2}}{(68.4043\times10^\mathrm{-9})^\mathrm{2}}}{\Hz^2} \\   
    ${ds_\mathrm{1}}/{d{T}}$ & \qty{5.31\times10^\mathrm{27}}{\Hz^2\per\K} & \qty{\pm8.96\times10^\mathrm{27}}{\Hz^2\per\K} \\
    $A_\mathrm{2}$ & \qty{\frac{0.4079426}{(116.2414\times10^\mathrm{-9})^\mathrm{2}}({2\pi c})^\mathrm{2}}{\Hz^2}  \\
    $s_\mathrm{2}$ & \qty{\frac{({2\pi c})^\mathrm{2}}{(116.2414\times10^\mathrm{-9})^\mathrm{2}}}{\Hz^2} \\   
    ${ds_\mathrm{2}}/{d{T}}$ & \qty{-2.23\times10^\mathrm{28}}{\Hz^2\per\K} & \qty{\pm5.17\times10^\mathrm{27}}{\Hz^2\per\K} \\
    $A_\mathrm{3}$ & \qty{\frac{0.8974794}{(9896.161\times10^\mathrm{-9})^\mathrm{2}}({2\pi c})^\mathrm{2}}{\Hz^2}  \\
    $s_\mathrm{3}$ & \qty{\frac{({2\pi c})^\mathrm{2}}{(9896.161\times10^\mathrm{-9})^\mathrm{2}}}{\Hz^2} \\   
    ${ds_\mathrm{3}}/{d{T}}$ & \qty{7.86\times10^\mathrm{25}}{\Hz^2\per\K} & \qty{\pm 1.42\times10^\mathrm{25}}{\Hz^2\per\K} \\
    \hline
  \end{tabular}
\end{table}

The fitted coefficients for \ce{SiO2} possess the same characteristics as \ce{Si3N4}, where a negative coefficient is only yielded for the resonance term closest to the experimentally surveyed frequencies. Here, the the closest resonance term ($j=2$) captures the primary physical mechanism of the red-shift of the absorption frequency, while the remaining oscillator terms ($j=1, 3$) serve as effective adjustment parameters.

\section{Estimation of local device temperature when achieving SHG}
\label{supsec:SHG temp}

  The temperature shift required to reach the experimentally demonstrated SHG within the microring resonator is estimated by considering two distinct thermal contributions. First, we calculate the temperature shift provided by the integrated heater as the applied voltage is increased. Using the $\mathrm{d}\nu/\mathrm{d}T$ value extracted from our comprehensive dispersive TRC model, a measured resonance frequency shift of $\approx$\qty{173.70}{\GHz} near \qty{193}{\THz} corresponds to a temperature increase of $\approx$\qty{31.02}{\K}. Second, we estimate the additional heating arising from laser power absorption. The circulating power within the microring resonator ($P_\mathrm{circ}$) is determined using the following expression~\cite{little_microring_1997}: 
  \begin{align}
    P_\mathrm{circ} = P_\mathrm{in} \cdot \frac{\lambda Q}{2\pi^\mathrm{2}Rn_\mathrm{eff}} 
  \end{align}
  where $P_\mathrm{in}$ is the input power of \qty{0.7}{\W} to \qty{1}{\W}, $\lambda = $ \qty{1554.25}{\nm} is the laser wavelength, $Q \approx 1 \cdot 10^6$ is the loaded quality factor of the resonance mode, \qty{R=22.85}{\um} is the ring radius, and $n_\mathrm{eff} \approx 2.1$ is the effective refractive index. With a high-quality \ce{Si3N4} waveguide the loss is on the order of \qty{0.1}{\dB\per\cm}, which is on par with that reported in prior literature~\cite{blumenthal_silicon_2018}, where $\approx$ \qty{20}{\percent} of the lost power is absorbed by the device~\cite{pfeiffer_ultra-smooth_2018, corato-zanarella_absorption_2024}. This results in roughly \qty{35}{\mW} to \qty{50}{\mW} of absorbed power. Applying a thermal resistance of approximately \qty{4.5}{\K\per\mW}~\cite{li_stably_2017}, we calculate a temperature shift of \qty{158}{\K} to \qty{225}{\K} due to laser absorption. Combining the integrated heater and laser absorption effects, the total local temperature shift required to achieve SHG is estimated to be between \qty{189}{\K} to \qty{256}{\K}.

\end{document}